\magnification = \magstep1

\hsize=7.0truein \vsize=9.4truein \hoffset=-0.3truein \voffset=-0.1truein

\nopagenumbers
    \def\cl{\centerline}
\def\vs{\vskip 5pt}  \def\ts{\thinspace}  \def\b{\kern -2pt}
\def\gapprox{$_>\atop{^\sim}$}  
\def\makeheadline{\vbox to 0pt{\vskip-24pt\line{\vbox to8.5pt{}\the
                  \headline}\vss}\nointerlineskip}
\def\toppageno{\headline={\hss\tenrm\folio\hss}}
\pretolerance=5000  \tolerance=5000
\def\etal{{\it et~al.~}} 
\def\nhi{\noindent \hangindent=15pt}

\headline={\leftskip = -0.3in
           \vbox to 0pt{PERSPECTIVES: ASTRONOMY (Submitted July 21, 2000 to
           {\it Science\/}) \hfill\null}}

\null \vskip -0.3truecm

\cl{\bf MONSTERS AT THE HEART OF GALAXY FORMATION$^1$} \vs

\cl{John Kormendy$^2$} \vs

    Black holes with masses of $10^6$\ts--\ts$10^{9.5}$ solar masses ($M_\odot$)
were invented in the 1960s to explain active galactic nuclei (AGNs) such as
quasars (1\ts--\ts4).  These supermassive black holes (BHs) stand in sharp
contrast to ordinary black holes with masses of a few $M_\odot$ that are well
known to form when massive stars die.  Supermassive black holes are more
mysterious.  Their origin is unknown, and their existence was a hypothesis.  But
it was a very successful hypothesis, and by the mid-1980s, an enormously
complicated AGN paradigm had been developed based on BH engines \hbox{(5, 6).}
The most immediate need then was to test whether dynamical evidence of BHs could
be found.  In the following decade, much effort was invested in looking for dark
objects in galactic nuclei (7, 8).  The evidence for them is now strong, and in
two objects -- our Galaxy and NGC 4258 -- the dark mass must live inside such a
small radius that astrophysically plausible alternatives to a BH can be excluded
(9).  Still, the emphasis was on checking and further developing the AGN
paradigm.  That is, the subject had not progressed much beyond its roots.  BHs
were studied mainly to understand the spectacular but restricted phenomena of
AGNs. 

      This situation is changing rapidly.  Surveys with the {\it Hubble Space
Telescope\/} (HST) are finding BHs in every galaxy that has an 
elliptical-galaxy-like ``bulge'' component (10).  This strengthens hints from
ground-based spectroscopy (11) that BHs are standard equipment in galaxy bulges.
The observations imply that BH growth and galaxy formation are closely linked.
The result is a profound change in how astronomers view BHs.  More than just
exotica needed to explain rare AGNs, they are becoming an integral part of our
understanding of galaxy formation. 

       This change in perspective was much in evidence in an all-day session on
BHs held on June 6, 2000, at the 196$^{\rm th}$ meeting of the American
Astronomical Society.  At least 15 new detections were reported, bringing the
total number of BHs available for study to at least 34.  

      The big news at the meeting was a new correlation between BH mass
$M_\bullet$ and host galaxy properties that was announced independently by Karl
Gebhardt (12, 13) and by Laura Ferrarese (14) and collaborators.  More massive
BHs live in galaxies whose stars have larger random velocities $\sigma$.  The
figure shows this correlation together with an older correlation (15, 7) between
BH mass and the total luminosity $L_{\rm bulge}$ -- surrogate for total mass -- 
of the bulge.  More massive BHs live in more massive bulges.  This is not
surprising: many properties of astronomical objects scale with mass.  In the
present case, more massive galaxies are expected to have more fuel to feed BHs.
But the scatter in the $M_\bullet$ -- $L_{\rm bulge}$ correlation is
substantial, and a few galaxies stand out as having anomalously large or small
BHs.  Remarkably, the new correlation has essentially zero scatter.  The figure
shows all BH detections, but if the sample is restricted to the galaxies with
the most accurate mass measurements, then the scatter is consistent with the
error bars. 

      Tight correlations in astronomy have a history of leading to fundamental
advances.  In this case, the correlation implies a connection between galaxy
formation and the process that feeds BHs, building them up to their present
masses while making them shine as quasars.  To be accreted onto a
BH, fuel must be robbed of almost all of its angular momentum.  This
is difficult, so the process of ``feeding the monster'' is poorly understood
(16).  But tying BH growth to galaxy formation is useful progress.  This is what
the new correlation so nicely accomplishes.

\vskip 8pt
\hrule width \hsize
\vskip 6pt

$^1$Based on observations with the NASA/ESA {\it Hubble Space Telescope},
    obtained at the Space Telescope Science Institute, operated by
    AURA, Inc., under NASA contract NAS 5-26555

$^2$Department of Astronomy, University of Texas at Austin, RLM 15.308, 
    Austin, TX 78712, USA.  E-mail: kormendy@astro.as.utexas.edu;
    URL: http://chandra.as.utexas.edu/$\sim$kormendy

\eject

\pageno=2 \toppageno
 
      The argument is as follows (17).  The two correlations in the figure are
almost equivalent, because bulge mass is proportional to $\sigma^2$.
But the $M_\bullet$ -- $\sigma$ correlation contains new information.  BH mass
is so closely connected with this new information that it fixes up the imperfect
correlation with total mass and produces a correlation with $\sigma$ that has
almost no scatter.  What is this new information?   Why, when a BH is unusually
massive for its luminosity, is it also high in $\sigma$, so that exceptions in
the left panel of the figure are not exceptions in the right panel?  There are
several possibilities.  The stellar mass-to-light ratio could be anomalously
large; by the virial theorem, equilibrium demands more velocity for more mass.
This proves not to be the main effect.  Instead, bulges with unusually high
velocity dispersions are unusually compact: they have higher surface
brightnesses and smaller radii than normal for their luminosities.  Then the
stars are closer together, so their gravitational forces on each other are
larger, so they must move faster.  We conclude that, when a galaxy is hotter
than average, it underwent more dissipation than average and collapsed inside
its dark halo to a smaller size and higher density than average.  But if BHs are
unusually massive whenever their bulges are unusually collapsed, then this
strongly suggests that BH masses were determined by the bulge formation process.

      In sharp contrast to their correlations with bulge properties, BHs
do not correlate with galaxy disks.  Pure disk galaxies -- ones that lack a
bulge component -- must have BH mass fractions that are much smaller
than the canonical 0.2\ts\% that is implied by the $M_\bullet$ -- $L_{\rm
bulge}$ correlation.  The best studied pure disk galaxy is our neighbor in the
Local Group of galaxies, Messier 33.  If disks contained BHs like bulges do,
then M\ts33 should have a black hole of mass $M_\bullet \sim 3 \times 10^7$
$M_\odot$.  But

\vfill

\includegraphics{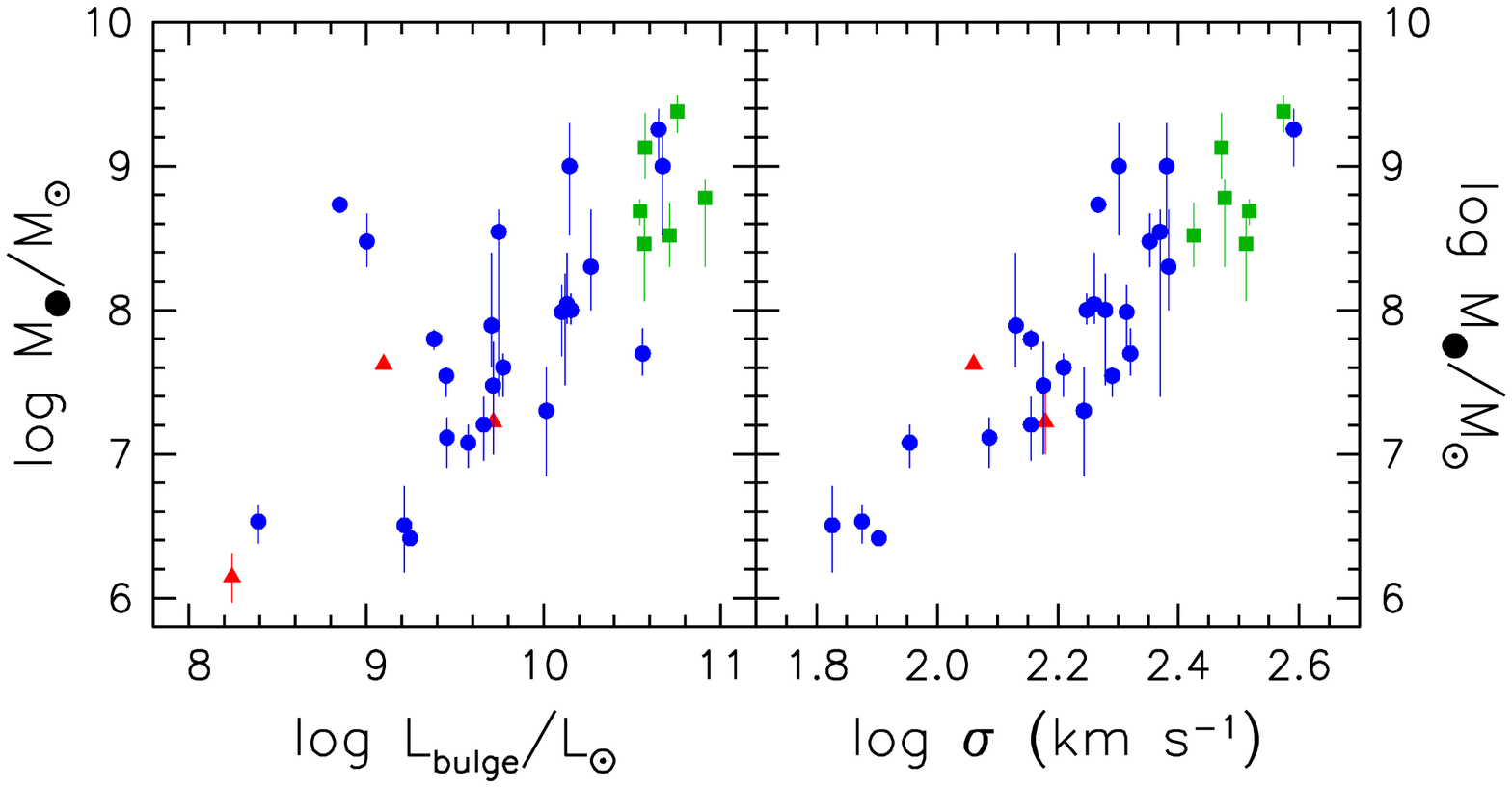}

\vskip 10pt

\noindent Correlations of BH mass with properties of the bulge component of the
host galaxy, i.{\ts}e., (left) total luminosity in units of the luminosity of
the sun and (right) mean velocity dispersion $\sigma$ inside the radius $r_e$
that contains half of the light of the bulge (13, 17).  It is important that the
radius used be large enough so that $\sigma$ is not inflated by the BH.  Apart
from this, the dispersion could be averaged in radii that are substantially
larger or smaller than $r_e$ without affecting the results.  In both panels,
filled circles indicate $M_\bullet$ measurements based on stellar dynamics,
squares are based on ionized gas dynamics, and triangles are based on maser disk
dynamics.  Note that exceptions to the \hbox{$M_\bullet$ -- $L_{\rm bulge}$}
correlation are not exceptions to the $M_\bullet$ -- $\sigma$ correlation.

\eject

\noindent HST spectroscopy shows that it cannot contain a BH more massive than
2000 $M_\odot$.  So BH growth is connected only with the process that forms
bulges.

  Bulges and elliptical galaxies (i.{\ts}e., diskless bulges) form when galaxies
collide and merge in a hierarchically clustering universe.   Gravitational
violence during mergers mixes orbits and redistributes energies and angular
momenta; the result is that quiescently formed systems such as disks are turned
into ellipsoids.  At least in the nearby universe, gas dissipation is required
to produce the high densities observed in bulges out of the low densities in
disks.  The new results on BHs suggest that the major events that form a bulge
and the major growth phases of its BH -- when it shone as an AGN -- were the
same events.  The likely formation process is a series of dissipative mergers
that fuel both starbursts and AGN activity.

      The exact relationship between BH growth and bulge formation is not known.
Some authors have suggested that BH growth happens before galaxy formation.  For
example, Silk and Rees (18) explore the idea that giant black holes come first
and then regulate how much galaxy can form around them via the radiation
pressure and gas outflows from the AGN.  We also know that some BH mass is
accreted after galaxy formation, because we observe low-level AGN activity in
some old galaxies.  What fraction of the mass is accreted before, during, and
after galaxy formation is likely to be a subject of debate in the years to come.

    However, a clear observational guide is provided by our best examples in the
local universe of the formation of giant ellipticals.  These are ``ultraluminous
infrared galaxies'' or ULIRGs; i.{\ts}e., infrared-bright galaxies with
luminosities $L$ \gapprox \ts$10^{12}$ $L_\odot$.  They are known to be mergers
in progress that involve large-scale dissipative collapse.~Sanders \etal 
(19, 20) have suggested that ULIRGs are quasars in formation.  This idea led to
a decade-long debate about whether ULIRGs are powered by active nuclei or by
starbursts.  Observations now suggest that both sides are correct: about 2/3
of the energy comes from starbursts and about 1/3 comes from nuclear activity
(21, 22).  In addition, submillimeter observations are finding high-redshift
versions of ULIRGs from the quasar era (23).  Again, many show AGN activity.
ULIRG properties are entirely consistent with the suggestion that bulge
formation, BH growth, and quasar activity all happen together. 

      Black holes affect galaxy formation in other ways, too.  For example, some
ellipticals have ``cuspy cores'', i.{\ts}e., density distributions that break at
small radii from steep outer power laws to shallow inner power laws. Faber \etal
(24) suggest that these cores may be produced via the orbital decay of binary
BHs.  When two galaxies merge, their BHs form a binary once violent relaxation
has finished making an elliptical. After that, the BHs sink toward the center by
flinging stars away.  This reduces the stellar density and may produce a break
in the density profile.

      Another possible way to make cores involves energy feedback from AGNs
(18).  If BHs are fed by the same dissipative collapse that makes bulges, then
the resulting AGN is easily energetic enough to affect the gas that is trying to collapse toward the BH.  It may prevent enough collapse to produce the deficit
of stars that we see as the core.

      Neither process is well studied, and neither is known to be the correct
explanation of cores.  But both are examples of how BHs may be a necessary
ingredient in our understanding of galaxy formation.  The developing
interconnection between BHs and other work on galaxy formation is one reason 
why the combined picture is compelling.

      Further progress is likely to be rapid.  We have just entered one of the
major payoff periods of the {\it Hubble Space Telescope\/}.  The search for
supermassive black holes was always a major goal of the telescope.  Since
the 1997 installation of the Space Telescope Imaging Spectrograph (STIS), the
search has become much more efficient, because STIS samples light from a
one-dimensional slit rather than from a single aperture.  Many groups are
conducting BH surveys.  The next few years should produce more detections
than we have had in the past 15 years of the BH search. \vs

      I thank the Nuker team (D.~Richstone, PI) for many years of fruitful
collaboration and stimulating discussion of the above topics.  The Nuker team
is supported by HST data analysis funds through grants GO-02600.01-87A and
GO-07388.01-96A.  I also thank the STIS GTO team (BH PI Richard Green) for
communicating their results before publication. \vs

      {\bf References} \vs

\nhi 1.~Zel'dovich, Ya.~B.~1964, {\it Soviet Physics -- Doklady}, {\bf 9}, 195

\nhi 2.~Salpeter, E.~E.~1964, {\it Astrophys.~J.}, {\bf 140}, 796

\nhi 3.~Lynden-Bell, D.~1969, {\it Nature}, {\bf 223}, 690

\nhi 4.~Lynden-Bell, D.~1978, {\it Physica Scripta}, {\bf 17}, 185

\nhi 5.~Rees, M.~J.~1984, {\it Annu.~Rev.~Astr.~Astrophys.}, 22, 471

\nhi 6.~Begelman, M.~C., Blandford, R.~D., Rees, M.~J.~1984. {\it
        Rev.~Mod.~Phys.}, {\bf 56}, 255

\nhi 7.~Kormendy, J., Richstone, D.~1995, {\it Annu.~Rev.~Astr.~Astrophys.},
        {\bf 33}, 581

\nhi 8.~Richstone, D., \etal 1998, {\it Nature}, {\bf 395}, A14

\nhi 9.~Maoz, E.~1998, {\it Astrophys.~J.~(Letters)}, {\bf 494}, L181

\nhi 10.~Gebhardt, K., \etal 2000, in preparation

\nhi 11.~Magorrian, J., \etal 1998, {\it Astr.~J.}, {\bf 115}, 2285

\nhi 12.~Kormendy, J., Gebhardt, K., Richstone, D.~2000, {\it Bull.~A.~A.~S.},
         {\bf 32}, 702

\nhi 13.~Gebhardt, K., \etal 2000, {\it Astrophys.~J.~(Letters)}, in press
        (astro-ph/0006289)

\nhi 14.~Ferrarese, L., Merritt, D.~2000, {\it Astrophys.~J.~(Letters)}, 
         in press (astro-ph/0006053)

\nhi 15.~Kormendy, J.~1993, in {\it The Nearest Active Galaxies}, 
         ed.~J.~Beckman, L.~Colina, H.~Netzer (Madrid: Consejo Superior de
         Investigaciones Cient\'\i ficas), 197

\nhi 16.~Gunn, J.~E.~1979, in {\it Active Galactic Nuclei}, ed.~C.~Hazard, 
      S.~Mitton (Cambridge: Cambridge University Press), 213
 
\nhi 17.~Kormendy, J., \etal 2000, in preparation

\nhi 18.~Silk, J., Rees, M.~J.~1998, {\it Astr.~Ap.}, {\bf 331}, L1

\nhi 19.~Sanders, D.~B., \etal 1988a, {\it Astrophys.~J.}, {\bf 325}, 74 

\nhi 20.~Sanders, D.~B., \etal 1988b, {\it Astrophys.~J.~(Letters)}, 
         {\bf 328} L35 



\nhi 21.~Genzel, R., \etal 1998, {\it Astrophys.~J.}, 498, 579

\nhi 22.~Lutz, D., \etal 1998, {\it Astrophys.~J.~(Letters)}, 505, L103


\nhi 23.~Ivison, R.~J., \etal 2000, {\it Mon.~Not.~R.~Astr.~Soc.},
         in press (astro-ph/9911069)

\nhi 24.~Faber, S.~M., \etal 1997, {\it Astr.~J.}, {\bf 114}, 1771

\vfill\eject\end